\documentclass[nohyperref]{article}

\usepackage{microtype}
\usepackage{graphicx}
\usepackage{subfigure}
\usepackage{booktabs} 
\usepackage[english]{babel}
\usepackage{hyperref}

\usepackage[accepted]{icml2022}

\usepackage{amsmath}
\usepackage{amssymb}
\usepackage{mathtools}
\usepackage{amsthm}
\usepackage{enumitem}
\usepackage{float}

\usepackage[capitalize,noabbrev]{cleveref}

\theoremstyle{plain}

\theoremstyle{definition}

\theoremstyle{remark}

\usepackage[textsize=tiny]{todonotes}

\icmltitlerunning{Generating Diverse Vocal Bursts with StyleGAN2 and MEL-Spectrograms}

\begin{document}

\twocolumn[
\icmltitle{Generating Diverse Vocal Bursts with StyleGAN2 and MEL-Spectrograms}



\icmlsetsymbol{equal}{*}

\begin{icmlauthorlist}
\icmlauthor{Marco Jiralerspong}{xxx}
\icmlauthor{Gauthier Gidel}{xxx,yyy}
\end{icmlauthorlist}

\icmlaffiliation{xxx}{Mila and Université de Montréal, Montreal, Canada}
\icmlaffiliation{yyy}{Canada CIFAR AI Chair}

\icmlcorrespondingauthor{Marco Jiralerspong}{marco.jialerspong@mila.quebec.ca}
\icmlkeywords{Machine Learning, ICML}

\vskip 0.3in
]



\printAffiliationsAndNotice{}  

\begin{abstract}
We describe our approach for the generative emotional vocal burst task (ExVo Generate) of the ICML Expressive Vocalizations Competition. We train a conditional StyleGAN2 architecture on mel-spectrograms of preprocessed versions of the audio samples. The mel-spectrograms generated by the model are then inverted back to the audio domain. As a result, our generated samples substantially improve upon the baseline provided by the competition from a qualitative and quantitative perspective for all emotions. More precisely, even for our worst-performing emotion (awe), we obtain an FAD of $1.76$ compared to the baseline of $4.81$ (for reference, the FAD between the train/validation sets for awe is $0.776$). Code for this work can be found at \href{https://github.com/marcojira/stylegan3-melspectrograms}{https://github.com/marcojira/stylegan3-melspectrograms}
\end{abstract}

\section{Introduction}
The Emotion Generation Track (ExVo Generate) of the ICML Expressive Vocalizations Competition presents a unique opportunity to examine generative models and optimize their ability to create audio samples that resemble a novel dataset of non-verbal vocal bursts from \cite{Cowen2022HumeVB}. The hope is that the resulting models from this endeavor can provide exciting insights into the interplay between emotion and audio.  

We train a conditional variant of StyleGAN2 \cite{stylegan2} on various processed mel-spectrograms generated from the data and find substantial improvements in generated sample quality over the baseline described in \cite{exvo}.

\section{Related Work}
While research on mel-spectrograms as a means to go between the audio and the visual domain is extensive, most work focuses on mapping mel-spectrograms to audio (using GANs or other techniques) and not the unconditional generation of the mel-spectrograms themselves \cite{melgan, melspectro1, melspectro2, melspectro3}. By combining the methods above with an additional network that maps text sequences to mel-spectrograms, state-of-the-art voice synthesis is possible \cite{tan2021survey}.

\looseness=-1
However, as the first part of this process often depends on being provided a text sequence as input: unconditional generation of mel-spectrograms is as of yet still poorly understood. This area is crucial for the non-verbal domain, where we cannot provide complete text input to the model. Instead, the model must perform either completely unconditional or conditional generation with very little information (e.g., just being provided a specific emotion to generate). 

The closest work examining the unconditional generation of non-verbal audio samples is \citet{soundoflaughter}, that uses an MSGGAN to generate realistic laughter from a relatively small dataset of laughter samples. Conditional generation of laughter using additional information such as gender and age is also briefly explored, and interpolations between genders demonstrate the possibility of learning a meaningful latent space.

\looseness=-1
\vspace{-0.15in}
\paragraph{Our Contributions.} We further the work by \cite{soundoflaughter} extending the generation to a more complex challenge (larger dataset with multiple emotions) and move to use StyleGAN2, which has yielded substantial improvements at more traditional image generation tasks \cite{stylegan2ada}.

\section{Method}
At a high level, our model's audio generation process consists of 4 steps:
\setlist{nolistsep}
\begin{enumerate}[noitemsep]
    \item Preprocess the audio to improve sample quality (for all emotions).
    \item Convert the processed audio to square mel-spectrograms of dimension $128\times128$ or $256\times256$.
    \item Train StyleGAN2 on the mel-spectrograms and use it to generate new ones.
    \item Convert back the generated mel-spectrograms to audio.
\end{enumerate}

\subsection{Audio preprocessing}

\looseness=-1
To improve sample quality, training, and subsequent generation, we aim to create samples composed almost solely of the vocal burst. As we observe that many of the dataset's samples have varying noise levels, we first denoise the audio using the \texttt{noisereduce} \cite{denoisesoftware, denoisepaper} library, which uses spectral gating (i.e., a filtering threshold on the spectrogram of audio samples) to remove noise. 

Once the audio is denoised, we trim the silence at the start/end of each sample. As samples often have varying levels of silence at the start/end of recordings (usually related to the participant recording the sample), the trimming ensures that a larger proportion of the mel-spectrogram contains information on the vocal burst. 

\looseness=-1
Finally, we remove empty noise samples by performing a simple thresholding test (i.e., removing all samples where $\geq 98$\% of the log-scale mel-spectrogram has an amplitude under -$4.0$) and removing all samples of length $\leq 0.05$s after trimming). Combined, this eliminates 1047 samples.

\subsection{Mel-spectrogram conversion}
We transform the remaining processed samples to mel-spectrograms using \textit{torchaudio}'s \cite{torchaudio} mel-spectrogram conversion using filter values from \textit{librosa} \cite{librosa} for a more straightforward inversion. Specifically, we generate two sets of different resolutions (128 and 256) to experiment with the impact of higher resolutions on audio quality and the training process.

The number of mel filterbanks determines the resolution of the resulting mel-spectrogram. We use $128$ for the first set and $256$ for the second. While the higher resolution does allow for higher quality audio after inversion, we find that the improvement is not significant compared to the potential improvement in generation quality (i.e. great low resolution mel-spectrograms will sound better once inverted than good high resolution mel-spectrograms).

The hop length is the distance between windows where the Short-Time Fourier Transform (STFT) is applied. We use 256 for both and either pad (with a pixel value of $1\mathrm{e}{-6}$) or truncate the resulting samples, so they are of size 128x128 or 256x256. As most samples (after being trimmed) are between 1 and 3 seconds in length, this process ensures there is not excessive padding or truncation. However, a more principled approach could be used in the future to truly minimize padding/truncation.

\begin{figure}[ht]
\begin{center}
\centerline{\includegraphics[width=\columnwidth]{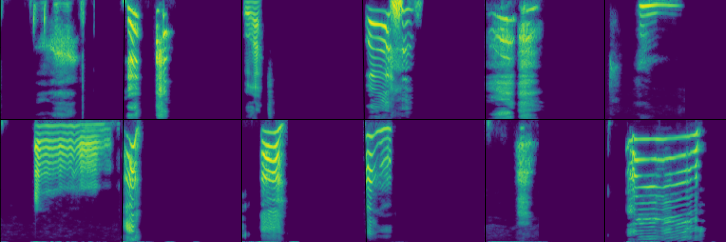}}
\caption{Examples of generated 128x128 mel-spectrograms using the processing steps described above on audio samples from the provided dataset. These samples have a single channel (i.e. are grayscale) but a Viridis colormap is used to make the data easier to digest. The figure contains 2 rows of 6 square samples.}
\label{real_samples}
\end{center}
\vskip -0.4in
\end{figure}

Lastly, we convert the samples to images by rescaling pixel values to be between 0 and 255. Using all of the above yields two datasets of grayscale samples (i.e. with one channel) of dimensions 128x128 or 256x256. Examples of the resulting samples can be seen in Figure \ref{real_samples}.

\subsection{StyleGAN training}
We use the StyleGAN2 architecture \cite{stylegan2ada} from the StyleGAN3 repository \cite{stylegan3} which takes advantage of an intermediate latent space that gets fed (after applying an affine transformation) to each layer of the generator through adaptive instance normalization. On the generated grayscale samples, we train both an unconditional model (on samples from all emotions at once) and a conditional one for both resolutions, yielding a total of 4 models. We set gamma to $.0512$ as recommended for images of that resolution and train for 72 hours with a batch size of 32 on a RTX 8000. 

The unconditional version is trained on the entire dataset (including the test set) while the conditional version is trained on the training + validation sets (as we do not have access to the classes of the test set). For each emotion, we select all samples whose score for that emotion is 1 (except for triumph which lacks samples).

\subsection{Mel-spectrogram inversion}
After generating mel-spectrograms using the trained StyleGAN2 models, we invert them back to audio samples. To do so, we begin by 
scaling them from [-1, 1] back to log scale and then reverse the transformation to log scale (i.e. $x = e^{2f(x)} - 1\mathrm{e}{-6}$). Finally, we apply \textit{librosa}'s \cite{librosa} mel-spectrogram inversion function, which makes use of the Griffin-Lim algorithm \cite{griffinlim} to convert mel-spectrograms to audio.

\section{Results}
The StyleGAN2 architecture learns to produce realistic mel-spectrograms in both the unconditional and conditional cases with a relatively little tuning. Audio samples from the best conditional model checkpoint are made available\footnote{\href{https://drive.google.com/drive/folders/1q9uw7dQhUUUZ3VR0AKwCFHyBrG1XW4AI?usp=sharing}{https://bit.ly/39Ea0gk}}. The FAD (Fréchet Audio Distance), HEEP (Human-Evaluated Expression Precision) and $S_{GEN}$ values of the best model can be found in Table \ref{sample-table}. Overall, we find that our StyleGAN2 based method significantly improves the generation of each emotion in terms of FAD and $S_{GEN}$ scores while HEEP is significantly improved for each emotion except Awe/Horror which are marginally worse.

Before continuing, we distinguish between visual FID \cite{fid} and the evaluation metric mentioned in \cite{exvo} which we designate as FAD (Fréchet Audio Distance), as it still examines the Fréchet distance of Gaussians but relies on the features from the last layer of a model trained on audio data instead of those from Inception Net~\citep{szegedy2016rethinking}. 

HEEP on the other hand uses human evaluations of the submitted samples where each sample is assigned a rating for each emotion based on how intensely the sample represents that emotion. The resulting matrix $H$ is then compared to the matrix $T$ of one-hot encoded samples (i.e. 1 for the emotion the sample is supposed to represent and 0 for the others), using:
\begin{equation}
    HEEP = \frac{\sigma_{TH}}{\sqrt{\sigma^2_T \sigma^2_H}}
\end{equation}

Finally, $S_{GEN}$ is an aggregate measure that combines FAD and HEEP:
\begin{equation}
    S_{GEN}{_e}={\frac {1/FAD_e+HEEP_e}{2}}.
\end{equation}

\begin{figure}[H]
\begin{center}
\centerline{\includegraphics[width=\columnwidth]{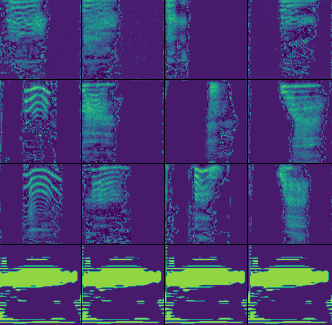}}
\caption{Progression of generated mel-spectrograms for the 128x128 conditional StyleGAN2. \textit{Top row:} Generator after four epochs. \textit{Second row:} Generator at lowest visual FID. \textit{Third row:} Best generator based on audio quality. \textit{Bottom row}: Beginning of mode collapse.}
\label{grid_progression}
\end{center}
\vskip -0.4in
\end{figure}

\begin{figure}[H]
\begin{center}
\centerline{\includegraphics[width=\columnwidth]{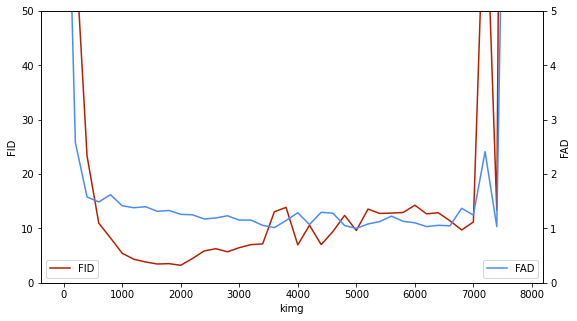}}
\caption{Comparison of FID and FAD progression for unconditional audio generation based on 128x128 mel-spectrograms. Interestingly, FID bottoms out quickly and then starts steadily increasing, whereas FAD decreases steadily (until training collapses).}
\label{fig:comparison_FID_FAD}
\end{center}
\vskip -0.2in
\end{figure}

\looseness-1
Figure~\ref{fig:comparison_FID_FAD} compares FAD and FID as training progresses. Even after only a few epochs, the generator learns to produce mel-spectrograms that, to the untrained eye, resemble those obtained from the real samples. Then, for most configurations we experimented with, the lowest visual FID is achieved roughly halfway through training. Nonetheless, qualitatively, sample quality keeps increasing before eventually training breaks as we observe mode collapse for all models (similarly to the collapse observed in \cite{biggan}). Visually, this progression can be observed in Figure~\ref{grid_progression}.


\begin{table*}[t]
\centering
\caption{
FAD, HEEP and $S_{GEN}$ values for each emotion. For FAD, 1000 samples are generated for each emotion using the best model checkpoint. The activation statistics of these samples are compared with the activation statistics of the samples in the validation set for the corresponding emotion (provided by \cite{exvo}). The row \textsc{Lower bound} corresponds to the FAD between the train and the validation set. Since an ideal model could not generate better samples than the ones of the validation set these values give a lower-bound on the best FAD we can expect. HEEP and $S_{GEN}$ scores are provided by the competition's organizers and computed as described in \cite{exvo}.}
\label{sample-table}
\vskip 0.15in
\begin{center}
\begin{small}
\begin{sc}
\begin{tabular}{lccccccccc}
\toprule
Emotion  & Amuse. & Awe & Awkward. & Distress & Excite. & Fear & Horror & Sadness & Surprise \\
\midrule
\multicolumn{10}{c}{FAD (lower is better)} \\
\midrule
Lower Bound    & .634 & .776 & 1.20 & .866 & .697 & .649 & 1.25 & .992 & .341 \\
This Work   & 1.28 & 1.76 & 1.76 & 1.77 & 1.75 & 1.57 & 1.34 & 0.94 & 1.67 \\
Baseline &4.92&  4.81 & 8.27 & 6.11 & 6.00 & 5.71 & 5.64 & 5.00 & 6.08   \\
\midrule
\multicolumn{10}{c}{HEEP (higher is better)} \\
\midrule
This Work  & 0.707 & 0.455 & 0.312 & 0.372 & 0.212 & 0.229 & 0.205 & 0.359 & 0.599 \\
Baseline & 0.490&  0.46 & 0.036 & 0.32 & 0.084 & 0.042 & 0.27 & -0.033 & 0.22   \\
\midrule
\multicolumn{10}{c}{$S_{GEN}$ (higher is better)} \\
\midrule

This Work & 0.744 & 0.512 & 0.440 & 0.467 & 0.392 & 0.433 & 0.476 & 0.711 & 0.598 \\
Baseline & 0.347&  0.334 & 0.078 & 0.242 & 0.125 & 0.109 & 0.224 & 0.084 & 0.192   \\

\bottomrule
\end{tabular}
\end{sc}
\end{small}
\end{center}
\vskip -0.1in
\end{table*}

To generate the samples provided for the competition, we ultimately select the conditional model at 128x128, sampled at a checkpoint slightly before mode collapse. We find, from manual inspection combined with FAD that it yields the highest quality samples. No truncation is used to improve the precision of generated samples. 

While we do not report results for the models trained on 256x256 mel-spectrograms we find that even though they achieve decent results, the 128x128 equivalent was almost always better in terms of sample quality. Nonetheless, the higher resolution models also suffered from the same issue of mode collapse.

\vspace{-0.05in}
\section{Discussion}

Generally, we find that StyleGAN2 is a suitable architecture for generating realistic mel-spectrograms as long as early-stopping or some manual equivalent is used with respect of FAD due to the presence of mode collapse.

\looseness-1
While StyleGAN2 has generally been more resistant to mode collapse than other architectures the issue in this case is perhaps exacerbated by a relatively high resolution of images combined with a small dataset. Further work is needed to better understand the cause and methods to avoid/delay it.

As shown in Figure \ref{fig:comparison_FID_FAD}, even after FID starts increasing, FAD keeps improving (as well as our qualitative evaluation of the generated audio samples). This discrepancy seems to indicate two things. FAD is a better metric for evaluating the generator than FID and StyleGAN2 is able to learn domain-specific features that are ordinarily not required for visual learning tasks (or at least these features are not learned by InceptionNet). The difference could also potentially explain why it is hard to visually observe the quality of the mel-spectrograms increasing (after the initial bit of training) even though the quality of the generated audio is still improving.

The experiments with 256x256 mel-spectrograms illustrate a tradeoff that needs to be made between improving the inversion quality (a higher resolution mel-spectrogram will be inverted to higher quality audio) and the difficulty of the image-generation task (it is harder to learn to generate higher resolution images). While 128x128 seemed to offer a good middle ground (a not too significant loss in inversion quality with a resolution that StyleGAN2 has generally performed well on), it is possible that a lower or higher resolution would better optimize that tradeoff.

\section{Conclusion}
While this initial foray into unconditional, non-verbal audio generation is promising, multiple parts of this audio generation process can be further optimized.

Due to time constraints, little work was done in regards to hyperparameter exploration (amount of preprocessing, mel-spectrogram generation parameters, StyleGAN2 training parameters, etc.). A more principled search of these hyperparameters (especially the resolution of the images) as well as the addition of some form of extra regularization could significantly improve performance. For example \cite{regularization} finds that a regularization term that penalizes the discriminator for having different outputs for real/fake images helps postpone mode collapse and stabilize training. Otherwise, recent work has been done on optimizer related techniques to stabilize training of GANs. Exploring other optimizers than Adam such as ExtraAdam \cite{gidel2020variational} could also address mode collapse.

Additionally, instead of using the Griffin-Lim algorithm for mel-spectrogram inversion, taking advantage of a model such as MelGAN \cite{melgan} could improve inversion quality and speed. It would be particularly interesting to see if the two sets of networks could perhaps be trained jointly or if it'd be better to train them separately (we could also potentially get better inversion from low resolution mel-spectrograms using MelGAN). 

While StyleGAN2 works well as a general purpose image generation framework, many of the optimizations it introduced are not designed with the domain of mel-spectrogram images in mind. Further domain-specific techniques (for example adapting the architecture to be able to generate rectangular images which are more suitable for audio) could bring about significant improvements in generation capabilities of mel-spectrograms.

Other than just improving generation capabilities, a principled investigation of the latent space of the trained GANs could yield interesting insights. It would be particularly relevant to attempt interpolation along different emotions. We could then perhaps answer questions such as: what sound is halfway between horror and awe?

\nocite{langley00}

\bibliography{example_paper}

\begin{thebibliography}{21}
\providecommand{\natexlab}[1]{#1}
\providecommand{\url}[1]{\texttt{#1}}
\expandafter\ifx\csname urlstyle\endcsname\relax
  \providecommand{\doi}[1]{doi: #1}\else
  \providecommand{\doi}{doi: \begingroup \urlstyle{rm}\Url}\fi

\bibitem[Afsar et~al.(2021)Afsar, Park, Paquette, Gidel, Mathewson, and
  Muller]{soundoflaughter}
Afsar, M.~M., Park, E., Paquette, E., Gidel, G., Mathewson, K.~W., and Muller,
  E.
\newblock Generating diverse realistic laughter for interactive art.
\newblock NeurIPS Workshop on ML for Creativity and Design, Sydney, Australia,
  2021.
\newblock \doi{10.48550/ARXIV.2111.03146}.

\bibitem[Baird et~al.(2022)Baird, Tzirakis, Gidel, Jiralerspong, Muller,
  Mathewson, Schuller, Cambria, Keltner, and Cowen]{exvo}
Baird, A., Tzirakis, P., Gidel, G., Jiralerspong, M., Muller, E.~B., Mathewson,
  K., Schuller, B., Cambria, E., Keltner, D., and Cowen, A.
\newblock {The ICML 2022 Expressive Vocalizations Workshop and Competition:
  Recognizing, Generating, and Personalizing Vocal Bursts}, 2022.

\bibitem[{B}rian {M}c{F}ee et~al.(2015){B}rian {M}c{F}ee, {C}olin {R}affel,
  {D}awen {L}iang, {D}aniel {P}.{W}.~{E}llis, {M}att {M}c{V}icar, {E}ric
  {B}attenberg, and {O}riol {N}ieto]{librosa}
{B}rian {M}c{F}ee, {C}olin {R}affel, {D}awen {L}iang, {D}aniel
  {P}.{W}.~{E}llis, {M}att {M}c{V}icar, {E}ric {B}attenberg, and {O}riol
  {N}ieto.
\newblock librosa: {A}udio and {M}usic {S}ignal {A}nalysis in {P}ython.
\newblock In {K}athryn {H}uff and {J}ames {B}ergstra (eds.),
  \emph{{P}roceedings of the 14th {P}ython in {S}cience {C}onference}, 2015.
\newblock \doi{10.25080/Majora-7b98e3ed-003}.

\bibitem[Brock et~al.(2019)Brock, Donahue, and Simonyan]{biggan}
Brock, A., Donahue, J., and Simonyan, K.
\newblock Large scale {GAN} training for high fidelity natural image synthesis.
\newblock In \emph{International Conference on Learning Representations}, 2019.

\bibitem[Cowen et~al.(2022)Cowen, Bard, Tzirakis, Opara, Kim, Brooks, and
  Metrick]{Cowen2022HumeVB}
Cowen, A., Bard, A., Tzirakis, P., Opara, M., Kim, L., Brooks, J., and Metrick,
  J.
\newblock The hume vocal burst competition dataset {(H-VB)} | raw data [exvo:
  updated 02.28.22] [data set].
\newblock \emph{Zenodo}, 2022.
\newblock URL \url{https://doi.org/10.5281/zenodo.6308780}.

\bibitem[Gidel et~al.(2019)Gidel, Berard, Vignoud, Vincent, and
  Lacoste-Julien]{gidel2020variational}
Gidel, G., Berard, H., Vignoud, G., Vincent, P., and Lacoste-Julien, S.
\newblock {A Variational Inequality Perspective on Generative Adversarial
  Networks}.
\newblock In \emph{International Conference on Learning Representations}, 2019.

\bibitem[Griffin \& Lim(1984)Griffin and Lim]{griffinlim}
Griffin, D. and Lim, J.
\newblock Signal estimation from modified short-time fourier transform.
\newblock \emph{IEEE Transactions on Acoustics, Speech, and Signal Processing},
  1984.
\newblock \doi{10.1109/TASSP.1984.1164317}.

\bibitem[Heusel et~al.(2017)Heusel, Ramsauer, Unterthiner, Nessler, and
  Hochreiter]{fid}
Heusel, M., Ramsauer, H., Unterthiner, T., Nessler, B., and Hochreiter, S.
\newblock {GAN}s trained by a two time-scale update rule converge to a local
  nash equilibrium.
\newblock In Guyon, I., Luxburg, U.~V., Bengio, S., Wallach, H., Fergus, R.,
  Vishwanathan, S., and Garnett, R. (eds.), \emph{Advances in Neural
  Information Processing Systems}, volume~30. Curran Associates, Inc., 2017.

\bibitem[Karras et~al.(2020{\natexlab{a}})Karras, Aittala, Hellsten, Laine,
  Lehtinen, and Aila]{stylegan2ada}
Karras, T., Aittala, M., Hellsten, J., Laine, S., Lehtinen, J., and Aila, T.
\newblock Training generative adversarial networks with limited data.
\newblock 2020{\natexlab{a}}.
\newblock \doi{10.48550/ARXIV.2006.06676}.

\bibitem[Karras et~al.(2020{\natexlab{b}})Karras, Laine, Aittala, Hellsten,
  Lehtinen, and Aila]{stylegan2}
Karras, T., Laine, S., Aittala, M., Hellsten, J., Lehtinen, J., and Aila, T.
\newblock Analyzing and improving the image quality of {StyleGAN}.
\newblock In \emph{Proceedings of the IEEE/CVF Conference on Computer Vision
  and Pattern Recognition (CVPR)}, June 2020{\natexlab{b}}.

\bibitem[Karras et~al.(2021)Karras, Aittala, Hellsten, Laine, Lehtinen, and
  Aila]{stylegan3}
Karras, T., Aittala, M., Hellsten, J., Laine, S., Lehtinen, J., and Aila, T.
\newblock Stylegan3.
\newblock \url{https://github.com/NVlabs/stylegan3}, 2021.

\bibitem[Kumar et~al.(2019)Kumar, Kumar, de~Boissiere, Gestin, Teoh, Sotelo,
  de~Br\'{e}bisson, Bengio, and Courville]{melgan}
Kumar, K., Kumar, R., de~Boissiere, T., Gestin, L., Teoh, W.~Z., Sotelo, J.,
  de~Br\'{e}bisson, A., Bengio, Y., and Courville, A.~C.
\newblock Melgan: Generative adversarial networks for conditional waveform
  synthesis.
\newblock In Wallach, H., Larochelle, H., Beygelzimer, A., d\textquotesingle
  Alch\'{e}-Buc, F., Fox, E., and Garnett, R. (eds.), \emph{Advances in Neural
  Information Processing Systems}, volume~32. Curran Associates, Inc., 2019.

\bibitem[Prenger et~al.(2019)Prenger, Valle, and Catanzaro]{melspectro2}
Prenger, R., Valle, R., and Catanzaro, B.
\newblock Waveglow: A flow-based generative network for speech synthesis.
\newblock In \emph{ICASSP 2019 - 2019 IEEE International Conference on
  Acoustics, Speech and Signal Processing (ICASSP)}, 2019.
\newblock \doi{10.1109/ICASSP.2019.8683143}.

\bibitem[Sainburg()]{denoisesoftware}
Sainburg, T.
\newblock timsainb/noisereduce: v1.0.
\newblock \doi{10.5281/zenodo.3243139}.

\bibitem[Sainburg et~al.(2020)Sainburg, Thielk, and Gentner]{denoisepaper}
Sainburg, T., Thielk, M., and Gentner, T.~Q.
\newblock Finding, visualizing, and quantifying latent structure across diverse
  animal vocal repertoires.
\newblock \emph{PLoS computational biology}, 16, 2020.

\bibitem[Shen et~al.(2018)Shen, Pang, Weiss, Schuster, Jaitly, Yang, Chen,
  Zhang, Wang, Skerrv-Ryan, Saurous, Agiomvrgiannakis, and Wu]{melspectro1}
Shen, J., Pang, R., Weiss, R.~J., Schuster, M., Jaitly, N., Yang, Z., Chen, Z.,
  Zhang, Y., Wang, Y., Skerrv-Ryan, R., Saurous, R.~A., Agiomvrgiannakis, Y.,
  and Wu, Y.
\newblock Natural tts synthesis by conditioning wavenet on mel spectrogram
  predictions.
\newblock In \emph{2018 IEEE International Conference on Acoustics, Speech and
  Signal Processing (ICASSP)}, 2018.
\newblock \doi{10.1109/ICASSP.2018.8461368}.

\bibitem[Szegedy et~al.(2016)Szegedy, Vanhoucke, Ioffe, Shlens, and
  Wojna]{szegedy2016rethinking}
Szegedy, C., Vanhoucke, V., Ioffe, S., Shlens, J., and Wojna, Z.
\newblock Rethinking the inception architecture for computer vision.
\newblock In \emph{Proceedings of the IEEE conference on computer vision and
  pattern recognition}, 2016.

\bibitem[Tan et~al.(2021)Tan, Qin, Soong, and Liu]{tan2021survey}
Tan, X., Qin, T., Soong, F., and Liu, T.-Y.
\newblock A survey on neural speech synthesis.
\newblock \emph{arXiv preprint arXiv:2106.15561}, 2021.

\bibitem[Tseng et~al.(2021)Tseng, Jiang, Liu, Yang, and Yang]{regularization}
Tseng, H.-Y., Jiang, L., Liu, C., Yang, M.-H., and Yang, W.
\newblock Regularizing generative adversarial networks under limited data.
\newblock In \emph{Proceedings of the IEEE/CVF Conference on Computer Vision
  and Pattern Recognition (CVPR)}, June 2021.

\bibitem[Wang et~al.(2018)Wang, Stanton, Zhang, Ryan, Battenberg, Shor, Xiao,
  Jia, Ren, and Saurous]{melspectro3}
Wang, Y., Stanton, D., Zhang, Y., Ryan, R.-S., Battenberg, E., Shor, J., Xiao,
  Y., Jia, Y., Ren, F., and Saurous, R.~A.
\newblock Style tokens: Unsupervised style modeling, control and transfer in
  end-to-end speech synthesis.
\newblock In Dy, J. and Krause, A. (eds.), \emph{Proceedings of the 35th
  International Conference on Machine Learning}, volume~80 of \emph{Proceedings
  of Machine Learning Research}. PMLR, 10--15 Jul 2018.

\bibitem[Yang et~al.(2021)Yang, Hira, Ni, Chourdia, Astafurov, Chen, Yeh,
  Puhrsch, Pollack, Genzel, Greenberg, Yang, Lian, Mahadeokar, Hwang, Chen,
  Goldsborough, Roy, Narenthiran, Watanabe, Chintala, Quenneville-Bélair, and
  Shi]{torchaudio}
Yang, Y.-Y., Hira, M., Ni, Z., Chourdia, A., Astafurov, A., Chen, C., Yeh,
  C.-F., Puhrsch, C., Pollack, D., Genzel, D., Greenberg, D., Yang, E.~Z.,
  Lian, J., Mahadeokar, J., Hwang, J., Chen, J., Goldsborough, P., Roy, P.,
  Narenthiran, S., Watanabe, S., Chintala, S., Quenneville-Bélair, V., and
  Shi, Y.
\newblock Torchaudio: Building blocks for audio and speech processing.
\newblock \emph{arXiv preprint arXiv:2110.15018}, 2021.

\end{thebibliography}
\bibliographystyle{icml2022}



\end{document}